\documentclass[journal]{IEEEtran}
\usepackage{amsmath}
\usepackage{amssymb}
\usepackage{booktabs}
\usepackage{xcolor}
\usepackage{subfigure}

\ifCLASSINFOpdf
\else
  \usepackage[dvips]{graphicx}
  \graphicspath{{../eps/}}
 \fi
\begin{document}
\title{Ghost Imaging Based on Recurrent Neural Network}
\author{Yuchen~He,
        Sihong~Duan,
        Jianxing~Li,
        Hui Chen,
        Huaibin~Zheng,
        Jianbin~Liu,
        Yu~Zhou,
        and~Zhuo~Xu
}


\maketitle

\begin{abstract}
Benefit from the promising features of second-order correlation, ghost imaging (GI) has received extensive attentions in recent years.
Simultaneously, GI is affected by the poor trade-off between sampling rate and imaging quality.
The traditional image reconstruction method in GI is to accumulate the action result of each speckle and the corresponding bucket signal.
We found that the image reconstruction process of GI is very similar to the Recurrent Neural Network (RNN), which is one of the deep learning algorithm.
In this paper, we proposed a novel method that effectively implements GI on the RNN architecture, called GI-RNN.
The state of each layer in RNN is determined by the output of the previous layer and the input of this layer, and the output of the network is the sum of all previous states.
Therefore, we take the speckle of each illumination and the corresponding bucket signal as the input of each layer, and the output of the network is the sum of all previous speckle and bucket signal, which is the image of the target.
The testing results show that the proposed method can achieve image reconstruction at a very low sampling rate (0.38$\%$).
Moreover, we compare GI-RNN with traditional GI algorithm and compressed sensing algorithm.
The results of different targets show that GI-RNN is 6.61 dB higher than compressed sensing algorithm and 12.58 dB higher than traditional GI algorithm on average.
In our view, the proposed method makes an important step to applications of GI.
\end{abstract}


\IEEEpeerreviewmaketitle

\section{Introduction}
\IEEEPARstart{G}{host} imaging (GI), which was first demonstrated by Pittman and Shih in 1995, is regarded as a novel imaging technology different than conventional methods based on first-order interference~\cite{pittman1995optical}.
By employing second-order correlation, GI provides some promising features such as lens-less imaging, turbulence-free and high detection sensitivity.
Consequently, GI has attracted much attention of researchers and produced many interesting results~\cite{valencia2005two, shapiro2008computational, meyers2008ghost, bromberg2009ghost, ferri2010differential, meyers2011turbulence, ryczkowski2016ghost, pelliccia2016experimental, khakimov2016ghost, Ota1246}.
However, GI is always trapped in the demand for a large number of speckles for high-quality imaging results, which has hindered the development of its applications.
Compressed sensing (CS) method has been considered to solve this bottleneck problem~\cite{katz2009compressive, katkovnik2012compressive, amann2013compressive, longzhen2014super, zhang2021computational}, but large computational cost limits the application of CS in GI.
Recently, artificial intelligence (AI) methods are more and more utilized to improve GI~\cite{2017Deep, shimobaba2018computational, he2018ghost, wang2018de, wang2019learning, bian2020a, wu2020sub, zhang2021ghost}.

In~\cite{2017Deep, shimobaba2018computational, he2018ghost}, the image obtained by traditional GI method is used as the input of the deep neural network, and the predicted by the network is the high resolution target image.
~\cite{wang2019learning} directly takes the bucket signal as the input of the network, and uses the multi-branch network with residual structure to reconstruct the target image.
In training process, the ground truth needs to be resized and binarized into a resolution of 32*32 binary image.
This method can reconstruct a clear image when the sampling rate is 6.25$\%$.
In~\cite{wu2020sub}, DAttNet structure is proposed to reconstruct the target image, and the structure is a network model similar to U-net.
In training process, the resolution of ground truth is 128*128 and the size of train set contains 200011 samples.
This method can reconstruct a clear image when the sampling rate is 5.45$\%$.
However, because the resolution of the ground truth of the network is 128*128, 893 detections are required to image a target even when the sampling rate is 5.45$\%$.

In this paper, we proposed an improved GI method based on recurrent neural network (RNN) called GI-RNN, which can image the target at a very low sampling rate.
Conventional neural networks can only individually process one input, and the previous input has nothing to do with the latter one.
On the other hand, RNN can deal with the issue when the previous input is associated with the latter.
In GI, the basic correlation method is the most widely used, which continuously accumulate the action results of speckle and bucket signal.
In this sense, the result of the basic correlation method is the correlation with all previous states, that is, the input of each state is associated with the input of the previous one.
In training, GI-RNN takes the speckle of each illumination and the corresponding bucket signal as the input of each state in RNN.
Each state consists of the output of the previous state (speckle and bucket signal at time t-1) and the input of the current state (speckle and bucket signal at time t).
In this way, the final output of RNN is the sum of all speckles and bucket signals, that is, the image of the target.
In testing, the fixed sequence of speckle in training is employed to illuminate unknown target.
To demonstrate the proposed method, we test it on MNIST at different sampling rates, and compare it with basic correlation algorithm and compressed sensing algorithm.
The testing results show that the proposed method can achieve image reconstruction when the sampling rate is only 0.38$\%$.
Moreover, the results of different targets show that GI-RNN is 6.61 dB higher than compressed sensing algorithm and 12.58 dB higher than traditional GI algorithm on average.

The rest of this paper is organized as follows.
Section \uppercase\expandafter{\romannumeral2} provides a brief survey of related work.
In Section \uppercase\expandafter{\romannumeral3}, a comprehensive introduction to the proposed method is provided.
In Section \uppercase\expandafter{\romannumeral4}, the proposed method is demonstrated and compared by extensive experiments.
The paper is concluded in Section \uppercase\expandafter{\romannumeral5}.

\section{Related Work}
\subsection{Ghost Imaging}
At the beginning, GI collects data through coincidence detection of two arms.
One arm uses a bucket detector without spatial resolution to receive the echo signal containing target information, and the other arm (without target) uses an array detector to collect the spatial information of the corresponding position.
Different from the traditional imaging technology which depends on intensity information, GI relies on intensity fluctuation information to achieve image reconstruction.
Until 2018, shapiro proposed computational ghost imaging (CGI) to obtain spatial information of corresponding position through calculation.
The emergence of CGI simplifies the two arms of GI into one arm, which improves its practicability.
CGI employs a sequence of random speckles to illuminate target, and detect the echo signal by a bucket detector without spatial resolution.
Then, the image reconstruction process can be expressed as follow called basic correlation method
\begin{equation}
T = \sum\limits_{i = 1}^N {{P_i} \cdot {B_i}},
\label{eq:gi}
\end{equation}
where ${P_i}$ represents the i-th speckle and ${B_i}$ represents the i-th bucket signal.
Through N times of illumination, target information T is obtained.

\subsection{Recurrent Neural Network}
RNN is a kind of neural network with sequence data as input, recursion in the evolution direction of the sequence, and all nodes are connected in a chain.
RNN is very effective for the data with sequence characteristics.
It can excavate the temporal information and semantic information in the data.
By using this ability of RNN, the deep learning model has made a breakthrough in solving the problems in natural language processing (NLP) fields such as speech recognition, language model, machine translation and temporal analysis.
The state of the system at time t can be expressed as
\begin{equation}
{h_t} = f\left( {{h_{t - 1}} + {x_t}} \right),
\label{eq:rnn}
\end{equation}
where ${h_t}$ represents the state of t and ${x_t}$ represents the input of t.
${h_{t - 1}}$ represents the state of the previous time.
Eq.~\ref{eq:rnn} shows that the state of t is determined by the output of the previous state t-1 and the input of t.

\section{The Proposed Method}
\subsection{Principle}
\begin{figure}[h]
\centering
\includegraphics[width = 8 cm]{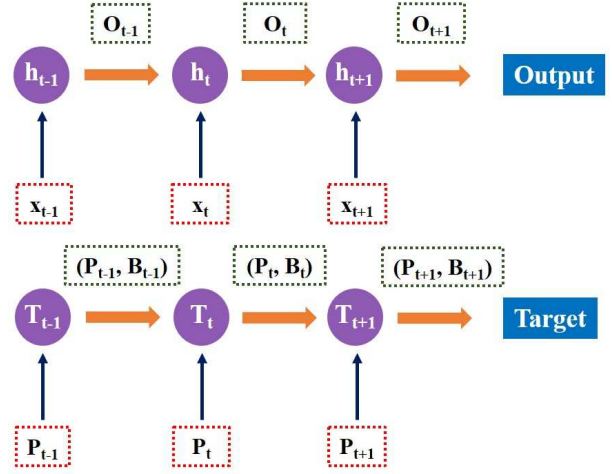}
\caption{Comparison of RNN and GI.}
\label{GIRNN}
\end{figure}

Inspired by the principle of basic correlation method and the process of RNN, we find that they are similar in form.
For GI, the basic correlation method combines the speckle of each illumination by the corresponding bucket signal, and accumulates the results.
For RNN, the output of the network is the sum of the previous states, and the state of each layer is determined by the output of the previous layer and the input of this layer.
Fig.~\ref{GIRNN} shows the corresponding relationship.

\begin{figure*}[!t]
\centering
\includegraphics[width = 13 cm]{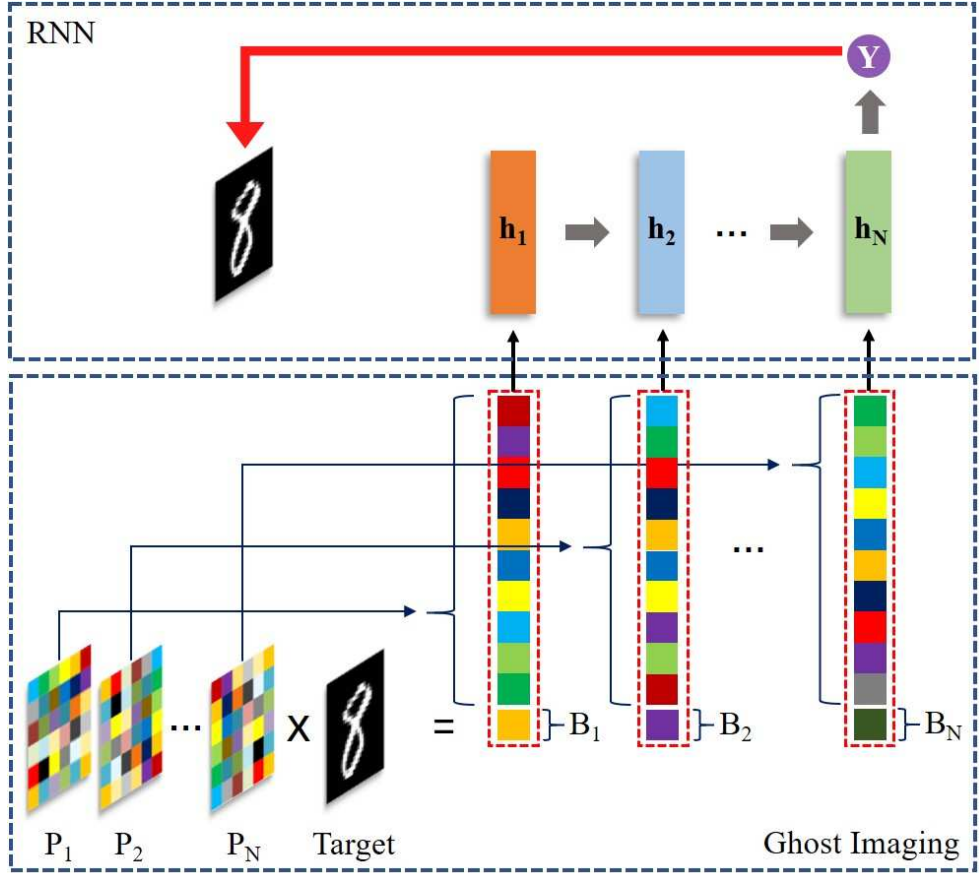}
\caption{Schematic diagram of the proposed method.}
\label{architecture}
\end{figure*}

In Fig.~\ref{GIRNN}, ${h_t}$  denotes the state of time t and ${T_t}$ denotes the target image of time t.
${x_t}$ denotes the input of time t and ${P_t}$ denotes the speckle pattern of time t.
${O_t}$ denotes the output of time t and ${({P_t},{B_t})}$ denotes the results of the interaction between speckle and bucket signal.
 As we can see, from expression to meaning, RNN is similar to GI.
Consequently, the action results of each speckle and bucket signal in GI can be regarded as the input of each layer in RNN.
Meanwhile, the imaging results of GI can be regarded as the output of the last layer of RNN.
The schematic diagram of the proposed method is shown in Fig.~\ref{architecture}.

\subsection{Architecture}
The proposed method consists of three parts: pre-processing module, RNN backbone and predictor.
The pre-processing module is used to convert speckle and corresponding bucket signal into a form suitable for RNN input.
In the proposed method, long short-term memory (LSTM) is used as the model of RNN.
One of the key reasons for using LSTM is that the network can effectively learn long-term dependent information, which is very consistent with the imaging process of GI.
GI-RNN is similar to the machine translation in NLP.
A speckle-bucket pair is equivalent to a "word", and the feature vector of the target is output after all the "words" are input.
When imaging complex targets, the data sequence will be relatively long, even the sampling rate is low.
Consequently, the network is required to have the ability to learn long-term dependent information, so LSTM is adopted.
In addition, LSTM can also effectively solve the problems of gradient vanishing and exploding in the process of long sequence training.
The predictor takes the feature vector output by LSTM as the input, and reconstructed image of target as the output.

\section{Experimental Results}
\subsection{Training Settings}
In this paper, we adopt the structure of multi-layer LSTM, and the number of circulating layers is 5.
The network input size is 785 and the hidden state size is 1024.
Moreover, the predictor input size is 1024 (equal to the LSTM hidden state) and output size is 784.
During training, we resize the target image as a one-dimensional vector with a length of 784 as the ground truth.
The mean-square error (MSE) is used as the loss function of network training between the reconstructed image and the ground truth, and we use Adam as the optimizer. After a large number of experiments, we found that the optimal initialization learning rate is 0.0001, and the weight decay is 0.
The train set in this work is MNIST with the image resolution of 28*28.
We randomly selected 9000 images from MNIST for training, and the testing samples are randomly selected from the testing set of MNIST.

\subsection{Results on MNIST}
\begin{figure}[!t]
\centering
\includegraphics[width = 9 cm]{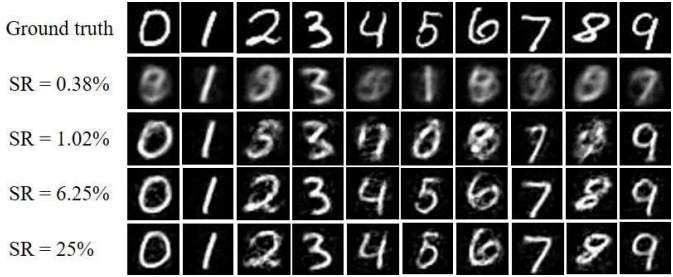}
\caption{Reconstruction results at different sampling rates.}
\label{rate}
\end{figure}

The experiments were compared at the sampling rates of 0.38$\%$, 1.02$\%$, 1.56$\%$, 2$\%$, 6.25$\%$, 25$\%$ and 100$\%$.
Then, we selected 0.38$\%$, 1.02$\%$, 6.25$\%$ and 25$\%$ for comparison.
The image reconstruction results are shown in Fig.~\ref{rate}.

\begin{figure}[!t]
\centering
\includegraphics[width = 17 cm]{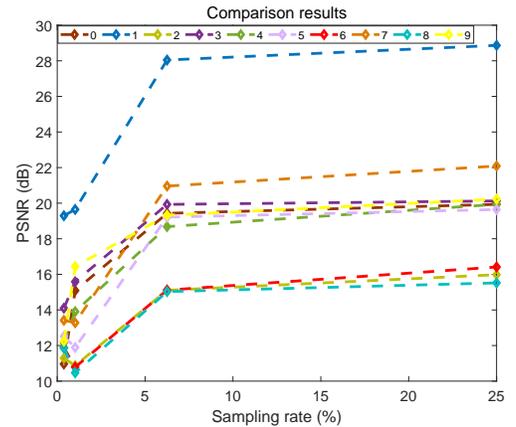}
\caption{PSNR comparison results under different sampling rates.}
\label{ratepsnr}
\end{figure}

As shown in Fig.~\ref{rate}, the proposed method can obtain stable results (the nine targets in the testing set) at the sampling rate of 6.25$\%$, and for some targets, the image can be obtained at 1.02$\%$ or even 0.38$\%$ ( which means 3 illuminations).
Fig.~\ref{ratepsnr} shows the result comparison curve of PNSR, and the detailed numerical results are shown in Table \uppercase\expandafter{\romannumeral1}.

\begin{table*}[!t]
\caption{PSNR comparison results under different sampling rates}
\label{table_sr}
\centering
\begin{tabular}{|l|l|l|l|l|l|l|l|l|l|l|}
\hline
            & Target\_0 & Target\_1 & Target\_2 & Target\_3 & Target\_4 & Target\_5 & Target\_6 & Target\_7 & Target\_8 & Target\_9 \\ \hline
SR = 0.38\% & 10.9614   & 14.9326   & 11.2939   & 14.0973   & 12.1563   & 12.5203   & 11.8820   & 13.4162   & 11.8558   & 12.2547   \\ \hline
SR = 1.02\% & 15.0871   & 23.0551   & 10.8728   & 15.5837   & 13.9068   & 11.8900   & 10.7861   & 13.2787   & 10.4665   & 16.4400   \\ \hline
SR = 6.25\% & 19.4287   & 27.0675   & 15.1048   & 19.9301   & 18.6813   & 19.2164   & 15.1146   & 20.9598   & 15.0306   & 19.3207   \\ \hline
SR = 25\%   & 19.9336   & 28.3169   & 15.9812   & 20.1249   & 19.9325   & 19.6416   & 16.4140   & 22.0872   & 15.5198   & 20.2247   \\ \hline
\end{tabular}
\end{table*}

\subsection{Comparison With Other Methods}
\begin{figure}[h]
\centering
\includegraphics[width = 8.5 cm]{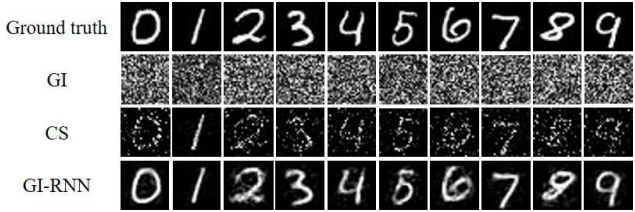}
\caption{Comparison of reconstruction results at sampling rate = 25$\%$.}
\label{Target}
\end{figure}

To demonstrate the performance of the proposed method, we compare GI-RNN with traditional GI and compressed sensing algorithm at sampling rate of 25$\%$ ( which means 196 illuminations).
The image reconstruction results are shown in Fig.~\ref{Target}.

\begin{figure}[h]
\centering
\includegraphics[width = 16 cm]{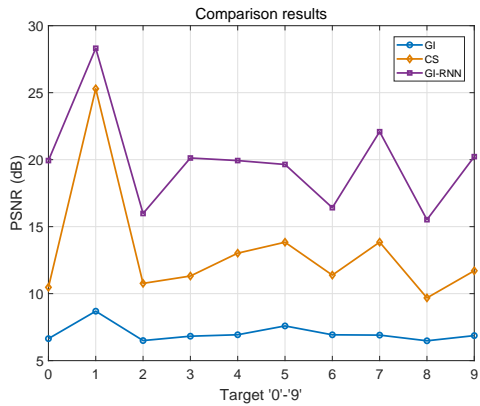}
\caption{PSNR comparison results of different methods.}
\label{PSNR}
\end{figure}

Fig.~\ref{Target} shows that the traditional GI algorithm (basic correlation) can not reconstruct target at such low sampling rate.
Compressed sensing algorithm (FISTA) can only partially reconstruct targets.
However, the proposed method can achieve stable results.
The PSNR curve of reconstruction results are shown in Fig.~\ref{PSNR}.

Fig.~\ref{PSNR} shows that the reconstruction results of GI-RNN are significantly better than the other two methods.
Table \uppercase\expandafter{\romannumeral2} lists the detailed numerical results.

\begin{table*}[!t]
\caption{PSNR comparison results of different methods}
\label{table_method}
\centering
\begin{tabular}{|c|c|c|c|c|c|c|c|c|c|c|}
\hline
       & Target\_0 & Target\_1 & Target\_2 & Target\_3 & Target\_4 & Target\_5 & Target\_6 & Target\_7 & Target\_8 & Target\_9 \\ \hline
GI     & 6.6399    & 8.6899    & 6.4968    & 6.8234    & 6.9307    & 7.5898    & 6.9267    & 6.9045    & 6.4805    & 6.8669    \\ \hline
CS     & 10.4739   & 25.2921   & 10.7656   & 11.3144   & 13.0200   & 13.8379   & 11.3820   & 13.8439   & 9.6817    & 11.7075   \\ \hline
GI-RNN & 19.9336   & 28.3169   & 15.9812   & 20.1249   & 19.9325   & 19.6416   & 16.4140   & 22.0872   & 15.5198   & 20.2247   \\ \hline
\end{tabular}
\end{table*}

After statistics (the nine targets in the testing set), Table \uppercase\expandafter{\romannumeral2} shows that the proposed method is 6.61 dB higher than compressed sensing algorithm and 12.58 dB higher than traditional GI algorithm on average.

\section{Conclusion}
In this paper, we propose a novel GI method based on RNN called GI-RNN.
The proposed method effectively integrates the traditional GI algorithm (basic correlation) into the architecture of RNN network, and can realize image reconstruction at a very low sampling rate.
The process of basic correlation method is to accumulate the results of each speckle interacting with the corresponding bucket signal.
For RNN, the state of each layer is determined by the input of this layer and the output of the previous layer, and the final output is determined by the accumulation of all previous layers.
Therefore, we take each illuminated speckle and corresponding bucket signal in GI as the input of each layer in RNN, and the final output of RNN is target image.
We demonstrate the proposed method at different sampling rates, and it can achieve image reconstruction when the sampling rate is only 0.38$\%$.
Moreover, we compare the proposed method with basic correlation method and compressed sensing method at sampling rate of 25$\%$.
Extensive experiments show that the proposed method is 6.61 dB higher than compressed sensing algorithm and 12.58 dB higher than traditional GI algorithm on average.

\ifCLASSOPTIONcaptionsoff
  \newpage
\fi

\bibliographystyle{IEEEtran}
\bibliography{RNN}

\end{document}